\documentclass[aps,prb,preprint,amsmath,amssymb]{revtex4}

\usepackage{epsfig}

\newcommand{\be}{\begin{eqnarray}}
\newcommand{\ee}{\end{eqnarray}}

\def\hatrho{{\hat\rho}}
\def\hatH{{\hat H}}
\def\hata{{\hat a}}

\begin{document}
\title{How Hot Is Radiation?}
\author{Christopher Essex}
\affiliation{Department of Applied
Mathematics, University of Western Ontario,
London, Ontario, Canada N6A 5B7}
\email{essex@uwo.ca}

\author{Dallas C. Kennedy}
\affiliation{The MathWorks, Inc., 3 Apple Hill
Drive, Natick, Massachusetts 01760}
\email{dkennedy@mathworks.com}

\author{R. Stephen Berry}
\affiliation{Department of Chemistry and James
Franck Institute, University of Chicago, Chicago, Illinois 60637}
\email{berry@uchicago.edu}

\begin{abstract}
A self-consistent approach to nonequilibrium radiation temperature is
introduced using the distribution of the energy over states. We
begin rigorously with ensembles of Hilbert spaces and end with
practical examples based mainly on the far from equilibrium
radiation of lasers. We show that very high, but not infinite, laser
radiation temperatures depend on intensity and frequency. Heuristic
``temperatures'' derived from a misapplication of equilibrium
arguments are shown to be incorrect. More general conditions for
the validity of nonequilibrium temperatures are also established.
\\
\centerline{Original: August 26, 2002}
\centerline{Final: June 8, 2003}
\centerline{Published in {\bf American Journal of Physics}}
\centerline{Vol. 71, No. 10, October 2003, pp. 969-978}
\centerline{PACS numbers: 05.70.Ln, 42.50.-p, 42.50.Ar, 42.55.Ah}
\end{abstract}

\maketitle

\section{Introduction}
\label{intro}

The standard definitions of intensive thermodynamic parameters, such as
temperature, seem to require the system in question to be in 
thermodynamic equilibrium. In this paper, we explore the entropy and
temperature of radiation out of equilibrium and show that, within limited
restrictions which do not require equilibrium, the radiation temperature
is well-defined and distinct from any associated matter temperature and
from plausible but incorrectly applied equilibrium definitions. We use
laser radiation as our major example.

Laser radiation is a fascinating example of a highly organized quantum
system of quasi-coherent bosons.\cite{sargent74,silf96,mandel95}
A laser beam is supported by external pumping, which keeps the beam far
from thermodynamic equilibrium. The laser shares this feature with other
steady-state systems that are kept from equilibrating by external
constraints. We find that the temperature of laser radiation far exceeds
the temperatures of the laser cavity and the lasing atomic transition.
Misidentifying the radiation with the matter temperature leads to
erroneous estimates of the laser radiation temperature that are as much
as ten orders of magnitude too small.

Photon number, unlike energy, is not conserved. The Gibbs-Duhem
relation for radiation, $SdT - VdP = 0$, implies that the two intensive
thermodynamic parameters, pressure $P$ (conjugate to volume) and
temperature $T$ (conjugate to energy), reduce to one independent intensive
parameter, which is usually identified as $T$.
This feature of radiation thermodynamics, like the photon's zero mass
and lack of rest frame, makes radiation thermodynamics much simpler than
that of matter, which has conserved particle numbers and nonzero chemical
potentials.\cite{essex99,reichl98} It also makes generalizing intensive
thermodynamic parameters out of equilibrium much easier. Thus radiation
is a natural context in which to introduce nonequilibrium temperature.

A properly defined nonequilibrium temperature has physical meaning.  It
occurs in the
{\it entropy production rate} $\Sigma$, an important measure of both
how far a system is from equilibrium and how fast it is approaching
equilibrium.\cite{degroot62} $\Sigma$ has a generic form rooted in the
equilibrium expression for the entropy differential, $dS = dQ/T$, where
$dQ$ is the differential of heat or random energy, the change in system
energy while holding volume and particle number constant.  The form
$\Sigma\sim J_Q(12)(1/T_1 - 1/T_2)$ expresses the entropy produced by two
subsystems (1,2) at temperatures $T_1, T_2$ as they exchange a heat flux
$J_Q(12)$. Subsystem temperatures occur naturally in expressions for
entropy production.
$\Sigma$ is positive semidefinite and vanishes if and only if
$T_1 = T_2$, the condition for thermal equilibrium. The heat flux
$J_Q(12)$ vanishes in this case as well.

The difference $1/T_1 - 1/T_2$ of the inverse temperatures is a measure
of how far out of equilibrium the two subsystems are. The heat flux
$J_Q(12)$ is a measure of how fast the subsystems are approaching
equilibrium with one another, assuming no external pumping of the system.
(With external pumping,
$J_Q(12)$ is a measure of how much power has to be injected into the
system to keep it from equilibrating.) The product of these two
quantities, given by $\Sigma$, combines the two measures into a single
quantity characteristic of a nonequilibrium process.

The books by Reichl\cite{reichl98} and De Groot and Mazur\cite{degroot62}
explain in detail the significance and role of entropy production in
nonequilibrium matter systems. Section~\ref{otherphysconseq} and
Refs.~\onlinecite{essex99} and \onlinecite{bludman97} explore entropy
production in radiation and radiation-matter systems.

\section{Misapplications of Temperature Out of Equilibrium}
\label{misapp}

Many formulas for various physical quantities have units of temperature. 
These formulas frequently are defined as the temperatures that
equilibrium systems would have if the energy, entropy, number, etc.,
were all rearranged in some particular way. There are many ways to
rearrange a system into an equilibrium state. Consequently, many
definitions of these pseudo
temperatures are possible, but none represent a temperature of the
actual system, and the application of temperature to nonequilibrium
systems is ambiguous.\cite{sien93}

Consider a helium-neon laser with the transition line $\lambda_0 =
632.8$\,nm. A heuristic but incorrect equipartition argument which sets
the mean photon energy equal to $k_BT$ would associate this transition
with a temperature $k_B T_0 = hc/\lambda_0$ or $T_0 \sim 2\times
10^4$\,K. The quantity $T_0$ is the temperature that a black body would
have if the distribution of energy among the photons were contrived to
have an average photon energy corresponding to the photons in question.
But the rest of the distribution is far from a blackbody for any laser.
$T_0$ refers to a photon gas with a different distribution, energy,
and entropy. So $T_0$ has no direct physical meaning for laser radiation,
which is far from equilibrium with itself and with its matter
source.

Laser radiation often is idealized as having infinite temperature,
as lasers are interpreted as a source of pure work, although ultimately
this interpretation is unsatisfactory. It would put a powerful X-ray laser
on an equal footing with a pocket red diode laser pointer, powered by
watch batteries. Is there no difference in ``temperature''? There seems to
be none if both lasers are described by infinite temperature.

Other possible pseudo temperatures include the temperatures that a beam
would have if the same energy or the same entropy were
arranged differently, for example, in a black body distribution. Such
temperatures do not reflect the actual distribution of energy,
entropy, and photon number in the beam.
Different as these definitions are from each other, they would all
agree if the laser radiation were forced into thermodynamic equilibrium.

There are other nonequilibrium temperature definitions that reflect the
actual energy and entropy distributions. For example, objects in the
laboratory surrounding the laser have meaningful local temperatures.
The material
around a He-Ne laser is at room temperature $T_{\rm room}\simeq 300$\,K,
and the He-Ne gas is at $T_{\rm gas}\simeq 400$\,K.

The inverted populations $N_2$ and $N_1$ of the upper and lower laser
energy levels $E_2$ and $E_1$ are associated with a temperature $T_{21}$
through a formal Boltzmann distribution,
\begin{equation}
N_2/N_1 = \exp[ -(E_2 -E_1)/k_BT_{21}] ,
\end{equation}
which does not hold for all levels. Such a pseudo temperature can be
defined for any two levels. In this case, the definition and population
inversion imply a negative value for $T_{21}$.  But such a definition
lacks a thermodynamic justification.

Although an inverted atomic population is essential to lasing action,
there is no reason to attribute the temperature $T_{21}$ to the radiation
field, a separate entity with its own thermodynamics.
In Secs.~\ref{noneqthermo} and~\ref{radthermo} we will identify the
natural radiation temperature that represents its distinctive nature,
and that points to a more general definition in terms of an
energy distributed over states. Temperature emerges
not as a proxy for energy, but instead as a measure of how the energy
is organized among the various microstates. It possesses a rigorous
definition and plays a natural role in nonequilibrium systems that go
beyond {\it ad hoc} heuristic estimates.\cite{silf96}

\section{Dynamics of the Radiation Field}

We analyze the radiation field in a box and decompose it into plane wave
modes.\cite{sargent74,merz69} Each field mode of wavevector ${\bf k}$ and
given polarization fills space. (We ignore polarization
in this paper for simplicity, which does not change the generality of
the argument.) The fundamental mode variable is its complex amplitude
$\alpha_{\bf k} = a_{\bf k} e^{i\phi_{\bf k}}$, in terms of its modulus
$a_{\bf k}$ and phase $\phi_{\bf k}$.

Each mode is a linear harmonic oscillator and can be mapped onto a quantum
harmonic oscillator in terms of lowering (raising) operators ${\hat
a}_{\bf k}$ $({\hat a}^{\dagger}_{\bf k})$, satisfying $[{\hat a}_{\bf k},
{\hat a}^{\dagger}_{\bf k}] = {\hat 1}_{\bf k}$, for each mode. The state
of a mode lives in an infinite-dimensional Hilbert space. The occupation
number basis
$|n\rangle$ forms a complete orthonormal set of number operator
eigenvectors corresponding to different photon numbers: ${\hat n}_{\bf
k}|n_{\bf k}\rangle = {\hat a}^{\dagger}_{\bf k}{\hat a}_{\bf k}|n_{\bf
k}\rangle = n_{\bf k}|n_{\bf k}\rangle$, and $\sum_n |n\rangle\langle n| =
{\hat 1}$. The field Hamiltonian
$\hatH = \sum_{\bf k} h\nu ({\hat n}_{\bf k}+1/2)$, with $\nu = ck$. The
state of the whole field is the direct product of all its mode states and
lives in a field Hilbert space (Fock space).

An alternative basis is the set of coherent states $|\alpha_{\bf
k}\rangle
$, eigenstates of the lowering operator: ${\hat a}_{\bf k}|\alpha_{\bf
k}\rangle = \alpha_{\bf k}|\alpha_{\bf k}\rangle$, defined so that the
occupation number expectation $N_{\bf k} = \langle\alpha_{\bf k}|{\hat
a}^{\dagger}_{\bf k} {\hat a}_{\bf k}|\alpha_{\bf k}\rangle =
|\alpha_{\bf k}|^2 = a^2_{\bf k}$. Coherent states satisfy
\begin{equation}
\int\! \frac{d^2\alpha}{\pi} |\alpha\rangle\langle\alpha | = {\hat 1} 
, \quad
\langle\beta |\alpha\rangle = e^{-\frac{1}{2}|\alpha -\beta |^2 + i{\rm
Im}(\beta^*\alpha)} , \quad
\langle n|\alpha\rangle = \frac{\alpha^n}{\sqrt{n!}} e^{-\frac{|\alpha
|^2}{2}} .
\end{equation}
 The state $|\alpha_{\bf k}\rangle$ is the quantum analogue
of a single classical mode ${\bf k}$ with amplitude
$\alpha_{\bf k}$. Physical results can be obtained with either the
coherent state basis or the photon occupation number basis
$|n\rangle$.\cite{merz69}

\smallskip Problem 1: Use the definition of $\langle
n|\alpha\rangle$ to derive the identities above: the result for
$\langle\beta |\alpha\rangle$ and the integral identity for
the coherent states
$|\alpha\rangle$. Also show directly, by expanding $|\alpha\rangle$ in
$|n\rangle$, that
$\hata |\alpha\rangle = \alpha |\alpha\rangle$.

Thermodynamics requires a statistical description with an ensemble of
many Fock spaces. Each mode has an ensemble of amplitudes and phases,
and the ensemble of the whole field is the direct product of the mode
ensembles.\cite{reichl98}

Very high temperatures are to be expected in low entropy, high energy
bosonic systems far from thermodynamic equilibrium such as the laser. 
Such high temperatures are natural for bosons, because they lack an
exclusion principle. Ideally, a multiparticle bosonic system could be
driven to zero entropy by putting all particles in one single-particle
state, without changing the system's total energy. As shown in
Sec.~\ref{infinite}, if the total energy remains fixed, the temperature
of that state would be infinite.

\section{Nonequilibrium Thermodynamics}
\label{noneqthermo}

A system's statistical ensemble defines the system's entropy. Defining
the temperature of a system or subsystem necessitates the restriction
to cases where its entropy and energy have a functional relationship. A
physically meaningful definition of temperature should reflect the
system's actual state and properly reduce to equilibrium temperature.

\subsection{The Entropy in General}
\label{generalentropy}

To define the entropy, we introduce an ensemble of a large number
$M$ of copies of a system assumed to be made up of discrete, countable
``things.'' Physically, these things are the system's fundamental
degrees of freedom. Because of statistical fluctuations, each copy is
microscopically different and distinguishable from the others.

Consider the entropy of two system copies. Let $W_1$ be the overall
number of ways that the first group of things can be arranged, and
$W_2$ the corresponding number for the second group of things. The
overall number of ways $W$ that the two groups can be arranged is the
product $W = W_1\times W_2$. For the
entropy, we want a function of $W$ that matches the usual definition
for the special case of equilibrium. Therefore, this function of $W$ must
depend on the two combined subsystems by adding the entropies of the
separate subsystems. Such a function must be a linear function of the
logarithm of $W$. The definition
$S = k_B\ln W$, up to a multiplicative factor and an irrelevant additive
constant, is the only function that does so. Note that entropy should be
additive, but need not scale in a simple way with the size of the
system. For a nonequilibrium system, although subsystems contribute
additively to the whole, the subsystems do not have to be homogeneous
and contribute to the whole in a simple scaling fashion.

Each system copy has the same internal probabilities
$p_\sigma$ for being in any particular state $\sigma$. The number
of system copies in state $\sigma$ is $M p_\sigma =
m_\sigma$. Suppose that the ensemble has $m_1$ systems in state
$\sigma_1$, $m_2$ systems in state $\sigma_2$, etc. The number of
ways this ensemble can be created from the system copies is
\begin{equation}
W_M = M!/[m_1! m_2!\dots m_\sigma !\dots ] ,
\end{equation}
and the
entropy for each member of the ensemble is $S = k_B\ln (W_M)/M$.
For $M$ large, Stirling's approximation yields
\begin{equation}
 M
S = k_B [M\ln M - \sum_\sigma (m_\sigma \ln m_\sigma)] .
\end{equation}
(Stirling's approximation is $n!\sim n^n e^{-n}\sqrt{2\pi n}$ for $n$
large. We ignore additive constants in the entropy, as these are
independent of the system's thermodynamic state.)

Because $m_\sigma = M p_\sigma$,
\be
S & = & k_B [\ln M - \sum_\sigma (p_\sigma \ln M + p_\sigma \ln
p_\sigma)]\nonumber \\
 & = & k_B [(1 - \sum_\sigma p_\sigma)\ln M - \sum_\sigma p_\sigma \ln
p_\sigma ]\nonumber \\
 & = &-k_B \sum_\sigma p_\sigma \ln p_\sigma ,
\ee
where the last line follows because $\sum_\sigma p_\sigma = 1$ and the
ensemble's system copies are independent.

A thermodynamic system's statistical ensemble is equivalent to its
normalized, Hermitian density operator $\hatrho :$ ${\rm
Tr}(\hatrho) = \sum_\sigma p_\sigma = 1$. The $p_\sigma$ are the
eigenvalues of $\hatrho$, with eigenvectors
$|\sigma\rangle$. Any observable ${\cal O}$ has a statistical average
\begin{equation}
\langle{\cal O}\rangle = {\rm Tr}(\hatrho {\hat{\cal O}}) =
\sum_\sigma p_\sigma \langle\sigma |{\hat{\cal O}}|\sigma\rangle .
\end{equation}
The average of the operator $-k_B\ln\hatrho$ is the entropy, $S =
-k_B{\rm Tr}(\hatrho\ln\hatrho)$. The mean energy $E
= {\rm Tr}(\hatrho\hatH)$ is the average of the Hamiltonian
$\hatH$.\cite{reichl98}

\subsection{The Validity of Nonequilibrium Temperature}
\label{valid}

Out of equilibrium, the entropy $S$ lacks a clear functional
dependence on the total energy
$E$, and the definition of $T$ becomes ambiguous. However, an important
generalization of
$T$ is still possible, with certain restrictions. The crucial
requirement is that the whole system decompose into subsystems whose
entropies are functions of each subsystem's energy only. That is, the
nonequilibrium system must decompose into subsystems each with its own
equilibrium. Restriction to full equilibrium is unnecessary.

We write the complete system Hamiltonian as $\hatH = \sum_p\hatH_p +
\sum_{q>p}\hatH_{pq}$, where the first sum is over separate subsystems
and the second is over all interactions between subsystems. The full
$\hatrho$ evolves via the quantum Liouville equation $i\hbar
d\hatrho/dt = [\hatH ,\hatrho ]$. For a particular subsystem $p$ to
have well-defined thermodynamics, the necessary restrictions are the
following:

(1) The whole system's density operator $\hatrho$ must be
factorizable into independent subsystems: $\hatrho =
\otimes_p\hatrho_p$. This factorization implies that the total system
entropy
$S$ is a sum over subsystems: $S = \sum_p S_p$.

(2) The commutator $[\hatH_p,\hatrho_p]$ must be negligible or zero.
Then $\hatrho_p$ and $\hatH_p$ are simultaneously diagonalizable, and
a functional relationship $\hatrho_p = \hatrho_p(\hatH_p)$ is possible.
The subsystem's ensemble probabilities are $p_{p_\sigma} =
\langle\sigma|\hatrho_p(\hatH_p)|\sigma\rangle = p_{p_\sigma}(E_{p_\sigma})$.
That is, each eigenvalue $p_{p_\sigma}$ of $\hatrho_p$ is a function of the
corresponding eigenvalue $E_{p_\sigma}$ {\it only}. It is not a function
of the other energy eigenvalues $E_{p_\mu}$ of subsystem $p$ or of the
$\{ E_{q_\mu} \}$ of the other subsystems $q$. (If the
interactions
$\hatH_{pq}$ between subsystems are negligible, then the total system
energy $E$ is a sum over the
$E_p$'s and each subsystem is in a stationary state. But this
restriction is not necessary. For example, laser radiation
modes in the laser cavity are strongly coupled to external
pumping.)

These special conditions allow a functional
relationship between ensemble averages of entropy and energy for
each subsystem that does not depend on the specific probability
distribution $\{ p_\sigma\}$. That is,
$S_p=S_p(E_p)$, which leads to a natural temperature of subsystem $p$:
\begin{equation}
\label{tempderivative}
\frac{1}{T_p} = {{\partial S_p}\over{{\partial E_p}}} ,
\end{equation}
where $T_p$ is a rate of change between extensive quantities, as it
normally is in thermodynamics. Such a relationship exists for photons if
the subsystems are chosen over small enough ranges of energy. We note that
(i) the functional relationship
$S_p = S_p(E_p)$ is an equation of state. (ii) If $S_p$ is a function of
$E_q$, for
$q\neq$ $p$, the temperature $T_p$ is still defined, but is not a
function of subsystem $p$ alone. (iii) If many subsystems are in
contact with one another, their temperatures represent a basis for their
complete thermal equilibrium. For each pair of subsystems $(p, q)$,
$T_p = T_q = T$, where $T$ is the single temperature of the whole
system:
$1/T = \partial S/\partial E$.

A familiar example of such a subsystem temperature occurs in the
local thermodynamic equilibrium of the radiation field
typical of stellar interiors.\cite{bludman97,mihalas99} In this case,
periodic boundary conditions are applied in small local boxes at
positions ${\bf r}$. A different field for each Fourier mode ${\bf
k}$, ensemble, and temperature $T({\bf k})$ can be defined in each box.
(Very long wavelength modes not fitting into the small local boxes
must be ignored.) Thus the subsystem is a particular mode ${\bf
k}$ in a box at ${\bf r}$, and $T_p$ becomes a photon phase space
temperature $T({\bf r},{\bf k})$. At a point ${\bf r}$ in space, this 
brightness temperature of the photons is a function of mode frequency $\nu$
= $ck$ and direction ${\bf\hat k}$ and is conventionally
written as $T_{\nu}({\bf\hat k})$. In Secs.~\ref{radthermo} and
\ref{reallaser}, we show how this temperature emerges naturally as a
genuine thermodynamic one.

\subsection{Pure States and Infinite Temperature}
\label{infinite}

A pure state has zero entropy. One of the $p_\sigma$ (say $p_\tau$) is unity, while the others
vanish, and so $S = 0$. There are many ways to realize a pure state with
an idealized laser. The two simplest are a pure coherent state $\hatrho =
|\alpha\rangle\langle\alpha |$ or pure occupation number state $\hatrho
= |n\rangle\langle n|$.

\smallskip Problem 2: Generally, $\hatrho$ is a mixture
of projection operators over different states; for example, in
the number basis, $\hatrho = \sum p_{nm}|n\rangle\langle m|$.
But the ensemble of a pure state $|\psi\rangle$ is $\hatrho =
|\psi\rangle\langle\psi|$, a single projection operator, not
a mixture. Prove that the eigenvalues of a pure ensemble can
only be one or zero. Hint: A normalized projection operator ${\hat\Pi}$
= $|\psi\rangle\langle\psi |$, with $\langle\psi |\psi\rangle = 1$,
satisfies ${\hat\Pi}^2 = {\hat\Pi}$.  Then infer that the eigenvalues $p$
of a pure $\hatrho$ satisfy the formally identical equation, $p^2 = p$.
%

We define a single mode's ensemble $\{ p_\sigma\}$. The ensemble average
energy is given by
$E =
\sum_\sigma p_\sigma E_\sigma$, where $E_\sigma = \langle\sigma |\hatH
|\sigma\rangle$. The subsystem temperature is defined by a derivative,
\begin{equation}
\label{deriv}
\frac{1}{k_BT} =
-\frac{\sum_\sigma \ln p_\sigma\, dp_\sigma}{\sum_\sigma E_\sigma \,
dp_\sigma} ,
\end{equation}
where $\sum_\sigma dp_\sigma = 0$.

This definition of subsystem temperature is a simple restatement
of the thermal equilibrium of subsystems across ensemble members. This
argument does not require all members of the ensemble to participate,
but the probabilities are assumed to be normalized over the active
members of the ensemble, with the rest are ignored.
Equation~(\ref{deriv}) thus provides a legitimate equilibrated
temperature of only those ensemble members indicated by the summation.

Consider two cases based on a sequence of equilibrated subsystem
probability distributions with the temperature defined by
Eq.~(\ref{deriv}):

(1) In equilibrium across all ensemble members, the $p_\sigma$ are
very small, as the probability is spread over the entire ensemble. As the
mean ensemble energy $E$ falls, while maintaining equilibrium, the average
is dominated by the lowest-lying value of the set $\{ E_\sigma\}$ as the
lower and lower energy states are occupied. Then $\{p_\sigma\}$ becomes
dominated by one
$p_\tau\rightarrow 1$, while the others vanish. But in most systems, the
denominator in Eq.~(\ref{deriv}) vanishes faster than the slowly-changing
logarithm in the numerator. So in equilibrium, $T\rightarrow 0$ as $E$ and
$S\rightarrow 0$. The classic example is the Bose-Einstein
condensate.\cite{reichl98}

(2) In the nonequilibrium case of interest in this paper, the mean
ensemble energy $E$ is fixed. Now redistribute this energy and the
ensemble probability into ensembles restricted to successively smaller
subspaces of the Hilbert space. Equilibrate the energy on these
successively smaller ensembles. Equation~(\ref{deriv}) still holds on
these smaller ensembles, but $E$ and the $\{p_\sigma\}$ are distributed
over fewer states. One $p_\tau$ approaches unity and the others vanish.
The number of ensemble members declines toward one, the pure state limit,
yielding, in the limit of one possible subsystem in the ensemble, the
limit
\begin{equation}
\frac{1}{k_BT} \sim -\frac{\ln p_\tau}{E_\tau}\rightarrow 0 
\end{equation}
for the temperature.
So $T\rightarrow\infty$ for a pure state if the total energy is fixed
while the entropy vanishes.

These results show how temperature is related not only to the ensemble
energies
$E_\sigma $, but to the ensemble distribution $\{ p_\sigma\}$ as well.

The radiation field of a real laser is not in a pure state. Each mode
has a statistical ensemble of amplitudes or occupation numbers, with an
associated entropy. Thus, whether the field is given a classical or quantum
description, a real laser does not have infinite temperature, although
laser temperatures are often idealized as infinite.

\section{Radiation Entropy and Temperature}
\label{radthermo}

{}From this point, we will work with individual photon modes ${\bf k}$ and
drop mode labels where they are not needed. Statistical ensembles of radiation
can be described by density operators $\hatrho$.\cite{sargent74,mandel95,reichl98}
Although a real laser is not in a coherent state, the coherent state basis is
exceptionally useful, because it connects the radiation mode description in terms
of amplitude and phase with photon number and energy.

In principle, we could calculate the entropy $S$ and other
thermodynamic functions from $\hatrho$ and find $T$. But $S$ is
difficult to calculate for arbitrary $\hatrho$. We consider only the
special case where the radiation field's statistical ensemble is
independent of the phase $\phi$. For most lasers, the phase is fully
randomized. The mode energy $E = h\nu N$ is always phase-independent,
depending only on the ensemble's photon number expectation $N = {\rm
Tr}({\hat a}^{\dagger}{\hat a})$. Thus the phase does not affect $S$ or
$T$ in this restricted case. This restriction simplifies $S(N)$ to a
general nonequilibrium form which we can obtain by a simple counting
argument appropriate for bosons.\cite{fnote1}

\subsection{Coherent State Basis --- Random Field Phase}

The density operator can be expanded over coherent states, $\hatrho = \!
\int \! d^2\alpha P(\alpha) |\alpha\rangle\langle\alpha |$, and then
projected on to the $|n\rangle$ basis, $p_{nm} = \langle n|\hatrho |m\rangle$.
The normalization is fixed,
${\rm Tr}(\hatrho) = \int \! d^2\alpha\ P(\alpha) =
\sum_n p_{nn} = 1$.\cite{fnote2}

The entropy is given by $S = -k_B{\rm Tr}(\hatrho\ln\hatrho) = -k_B\sum_n
\langle n|\hatrho\ln\hatrho |n\rangle$. We restrict the distribution
function
$P(\alpha)$ to be independent of phase $\phi$ and depend only on the
modulus $a$. The matrix $p_{nm}$ then becomes diagonal:
\begin{equation}
p_n\equiv p_{nm} = \!\int \! d^2\alpha \langle
n|\alpha\rangle\langle\alpha |m\rangle P(\alpha) =
2\pi\delta_{nm}\int^{\infty}_0 da
\frac{a^{2n+1}}{n!} e^{-a^2} P(a) .
\end{equation}
The entropy simplifies to $S = -k_B\sum_n p_n\ln p_n$. The eigenstates
$|\sigma\rangle$ of $\hatrho$ are the occupation number states
$|n\rangle$. The distribution $P(\alpha)$ is phase-independent if the
phase distribution for the amplitude $\alpha$ is random over the
interval $\phi\in [0, 2\pi)$.

The restriction to random phase and the result that the mode state can
be characterized by counting the photon number alone validates the
phase space approach to photon thermodynamics. For each mode of the
field labeled by a wavenumber ${\bf k}$, the mode state is determined
solely by its mean photon occupation number, $N_{\bf k} = \sum_n
p_{{\bf k}_n} n_{\bf k}$.\cite{fnote3}

\subsection{Phase Space Reduction: Counting Photons}
\label{counting}

Making the field's statistical ensemble phase-independent allows us to
find $S(N)$ by simply counting states for identical bosons, treating
the field quanta as particles. In this case, all of the entropy is due
to the randomness of the phase. The reasoning of this section parallels
that of Sec.~\ref{generalentropy}.

Consider $N$ identical things and $G$ identical possible places to put
them. Imagine not knowing where the $N$ things are among the $G$ places. 
Then there are $(N+G-1)!/N!(G-1)! = W$ ways of arranging the $N$ things
among the $G$ places. Now introduce an ensemble of $M$ system copies, with
each copy labeled by $\sigma$. For systems such as lasers with large numbers
of photons, we can assume that $G$ and $N\gg 1$ and use Stirling's approximation. 
$G$ and $N$ depend on the system copy $\sigma$ in the ensemble. The entropy is then:
\be
\label{entropy}
M S & = & k_B \sum_\sigma [ -G_\sigma \ln
G_\sigma + (G_\sigma + N_\sigma)\ln (G_\sigma + N_\sigma)-
N_\sigma \ln N_\sigma ]\nonumber \\
 & = & k_B \sum_\sigma G_\sigma [(1 + n)\ln (1+n) -n \ln n] ,
\ee
where $n_\sigma = N_\sigma /G_\sigma$ is the mean occupation number in the ensemble
and approaches the $\sigma$-independent limit $n = N/G,$ with $G = \sum_\sigma G_\sigma /M,$
as $M\rightarrow\infty$. Because $h\nu /T = \partial S/\partial N$ and $N = G n$,
\begin{equation}
\label{temperature}
k_BT = \frac{h\nu}{\ln(1+1/n)} .
\end{equation}

\smallskip Problem 3: Combine Eqs.~(\ref{entropy}) and
(\ref{temperature}) to express $S$ in terms of $T$.

To connect the result~(\ref{temperature}) to radiation observables, identify the
$G$ ``places'' with the Fourier space of modes ${\bf k}$ and use the general relation
of mean occupation number $n$ to the specific intensity $I_\nu ({\bf\hat k}):$
$n = c^2I_\nu ({\bf\hat k})/h\nu^3,$ where $I_\nu ({\bf\hat k})$ is the radiation
energy/area/time/frequency/solid angle.\cite{mihalas99,fnote4}
The resulting expression for the temperature,
\begin{equation}
\label{temperature/intensity}
k_BT_\nu ({\bf\hat k}) = 
\frac{h\nu}{\ln\bigl[ 1+h\nu^3/c^2I_\nu ({\bf\hat k})\bigr]} ,
\end{equation}
can also be derived from the Planck blackbody expressions
for $I_\nu$ by solving for $T$ and {\it defining} the resulting
temperature in terms of $I_\nu$.
The value of $I_\nu$ can then be arbitrary.\cite{essex99} The
radiation temperature varies as a function of beam direction ${\bf\hat
k}$ as well as frequency $\nu$. Except at zero frequency, $T_\nu$ is
zero only for vanishing $n$ or $I_\nu$ (photon vacuum).

\subsection{Temperature as a Lagrange Multiplier}
\label{lagrange}

Instead of Eq.~(\ref{deriv}) with its assumptions about probabilities,
the temperature can be considered as a Lagrange multiplier for holding
the ensemble mean energy fixed while maximizing the entropy.
This procedure can be extended to
subsystems, such as a single radiation mode or a collection of modes
constituting a Fourier subspace
$K$, a part of the whole mode space. The conclusions of the previous
sections are confirmed in a different way, as long as the mode phases are
random.

If we express Eq.~(\ref{entropy}) in Fourier space, the real space volume density
$s$ of entropy contributed by photons of wavenumber ${\bf k}$ and frequency
$\nu = ck$ in the Fourier subspace $K$ becomes:
\begin{equation}
\label{entropyDens}
s = \!\int_K \! k_B [ (1+n) \ln (1+n) - n \ln n]
\frac{d^3k}{(2\pi)^3} ,
\end{equation}
where $n$ is the mean occupation number. The real space volume density $e$
of energy in $K$ is
\begin{equation}
\label{energyDens}
e = \!\int_K \! nh\nu \frac{d^3k}{(2\pi)^3} .
\end{equation}
The maximum entropy at fixed energy is given by
\begin{equation}
\label{variation}
0 = \delta (s + \beta e) = \int_K \! \delta [\ln \Big({1+n \over
n}\Big) -
\beta h\nu] \frac{d^3k}{(2\pi)^3} ,
\end{equation}
from which we conclude that
\begin{equation}
\label{blackbn}
n = {1\over{e^{\beta h\nu} -1}} .
\end{equation}
We find an apparent black body distribution for the specific intensity:
\begin{equation}
\label{black}
I_\nu= {{h\nu^3}\over{c^2}} {1\over{e^{h\nu/k_BT} - 1}},
\end{equation}
and identify the Lagrange multiplier $\beta = 1/k_BT$. We define the
specific intensity of the entropy radiation
$J_\nu$ by photon counting, the same way that the specific energy
intensity
$I_\nu$ is defined.\cite{essex99} Then we find another version of 
Eq.~(\ref{tempderivative}):
\begin{equation}
\label{temptransport} {1 \over T_\nu (\mathbf{\hat{k}})} = { { \partial
J_\nu (\mathbf{\hat{k}}) } \over { \partial I_\nu (\mathbf{\hat{k}}) } } .
\end{equation}

If $K$ were all of Fourier space, this distribution would be an
equilibrium one with a single temperature $T$. But we have not
restricted $K$. If $K$ is any part of Fourier space, the distribution
is a black body distribution, but truncated to only its domain of
Fourier subspace. This subspace is complete with its own temperature
$k_BT[K] = \delta e[K]/\delta s[K]$, even though the whole distribution
is not present. The simplicity of radiation thermodynamics converts the
functions $s = s(n[K])$ and $e = e(n[K])$ into an implicit function
$s(e)$ over $K$, matching the more general argument of Sec.~\ref{valid}.
Therefore each $K$ forms a distinct thermodynamic system with a
temperature given by Eq.~(\ref{temperature/intensity}) or obtainable
by inverting Eq.~(\ref{black}).

This argument can be extended to a $K$ of zero volume. By squeezing a
fixed radiation energy into a vanishingly small frequency interval
$\Delta\nu$ and beam solid angle cone $\Delta\Omega
$, we recover the infinite temperature of a pure state for one mode ${\bf
k}$, with $\nu = ck$, as we now show.

Consider a fixed arbitrary specific intensity $I_\nu$ distribution and
break it into two parts, a black body function $B_\nu $ of constant
temperature
$T$, plus a finite deviation distribution $D_\nu({\bf\hat{k}})$, which
may depend on the direction ${\bf\hat{k}}$. Thus $I_\nu = B_\nu +
D_\nu$. The energy density $u$ within $(\Delta\nu , \Delta\Omega)$ is
given by
\be
\label{E}
cu & = & \! \int_{\Delta\Omega}\! d\Omega_{\bf k} \! \int_{\Delta\nu}
\! d\nu I_\nu
 = \!\int_{\Delta\Omega}\! d\Omega_{\bf k}\int_{\Delta\nu} d\nu (B_\nu
+ D_\nu)\nonumber \\
& = & \! \int_{\Delta\Omega}\! d\Omega_{\bf k}\int_{\Delta \nu} d\nu
{{h\nu^3/c^2}\over{e^{h\nu /k_BT} -1}}
+ \! \int_{\Delta\Omega}\! d\Omega_{\bf k}\!\int_{\Delta \nu}\!
d\nu\ D_\nu .
\ee
If we change to the variable $x = h\nu /k_BT$, the first integral of the
right-hand side of Eq.~(\ref{E}) can be recast as:
\begin{equation}
\label{integral}
{h\over{c^2}}\Big({{k_BT} \over h}\Big)^4 \! \int_{\Delta\Omega}
d\Omega_{\bf k}\!\int_{\Delta x}dx\ {x^3\over {e^{x} -1}} .
\end{equation}
It is clear that letting $\Delta\nu$ go to zero implies that $\Delta x$
goes to zero also. From Eq.~(\ref{temperature/intensity}) $T$ cannot
reach zero except for zero intensity, $I_\nu\rightarrow 0$. This limit
cannot be reached because we require that this beam have fixed, finite
energy density $u$. Because the integrand is finite, the double integral in
Eq.~(\ref{integral}) vanishes in the limit of
$\Delta\nu$ and $\Delta\Omega$ going to zero.

It follows that the second double integral on the right side of
Eq.~(\ref{E}) vanishes in this limit, as does the double integral in
Eq.~(\ref{integral}). If we use Eq.~(\ref{integral}) to rewrite
Eq.~(\ref{E}), we obtain
\begin{equation}
\label{limit}
\lim_{{\Delta\nu\rightarrow 0}\atop{\Delta\Omega\rightarrow 0}}
{h\over{c^2}}\Big({{k_BT}\over h}\Big)^4 =
 \lim\limits_{{\Delta\nu\rightarrow 0}\atop{\Delta\Omega\rightarrow 0}}
\bigg\{\Bigl(cu -
 \! \int_{\Delta\Omega}
\! d\Omega_{\bf k}\int_{\Delta \nu} d\nu
D_\nu\Big)
 \big[ \int_{\Delta\Omega} d\Omega_{\bf k}
 \!\int_{\Delta x}\! dx {x^3\over{e^x -1}}\big]^{-1} \bigg\} =
\infty .
\end{equation}
Thus $T$ becomes infinite when a fixed energy is concentrated
into a single mode.
This limit recovers the special ensemble of a pure state discussed in
Sec.~\ref{infinite}, fixed energy with zero entropy. The ensemble
degenerates into a single Hilbert space, with an exact photon number
$N_{\bf k} = E/h\nu$ or field mode modulus $a_{\bf k} =
\sqrt{E/h\nu}$. In this limit, these ensemble averages have no
statistical uncertainty. The mode fills all of real space and has one
wavelength and direction, a fixed absolute amplitude and energy density,
but random phase. The phase is independent of energy and does not affect
$T$. This result formally justifies the idealized picture of laser
radiation as having infinite temperature.

When radiation is in equilibrium with matter, all Fourier subspace temperatures
become the same finite value, returning $I_\nu$ to a full black body distribution,
as in Eq.~(\ref{black}), but valid over all modes. If the matter is selective in
its frequency or directional response to radiation, the relaxation of the radiation
to equilibrium is similarly limited. (The radiation intensity distribution can be found
in general by solving the equation of transfer that describes radiation transport
through matter.\cite{mihalas99})

\subsection{Temperature versus Intensity --- Classical Limits}
\label{classical}

\begin{figure}[h]
\centerline{\epsfxsize=6.6in \epsfbox{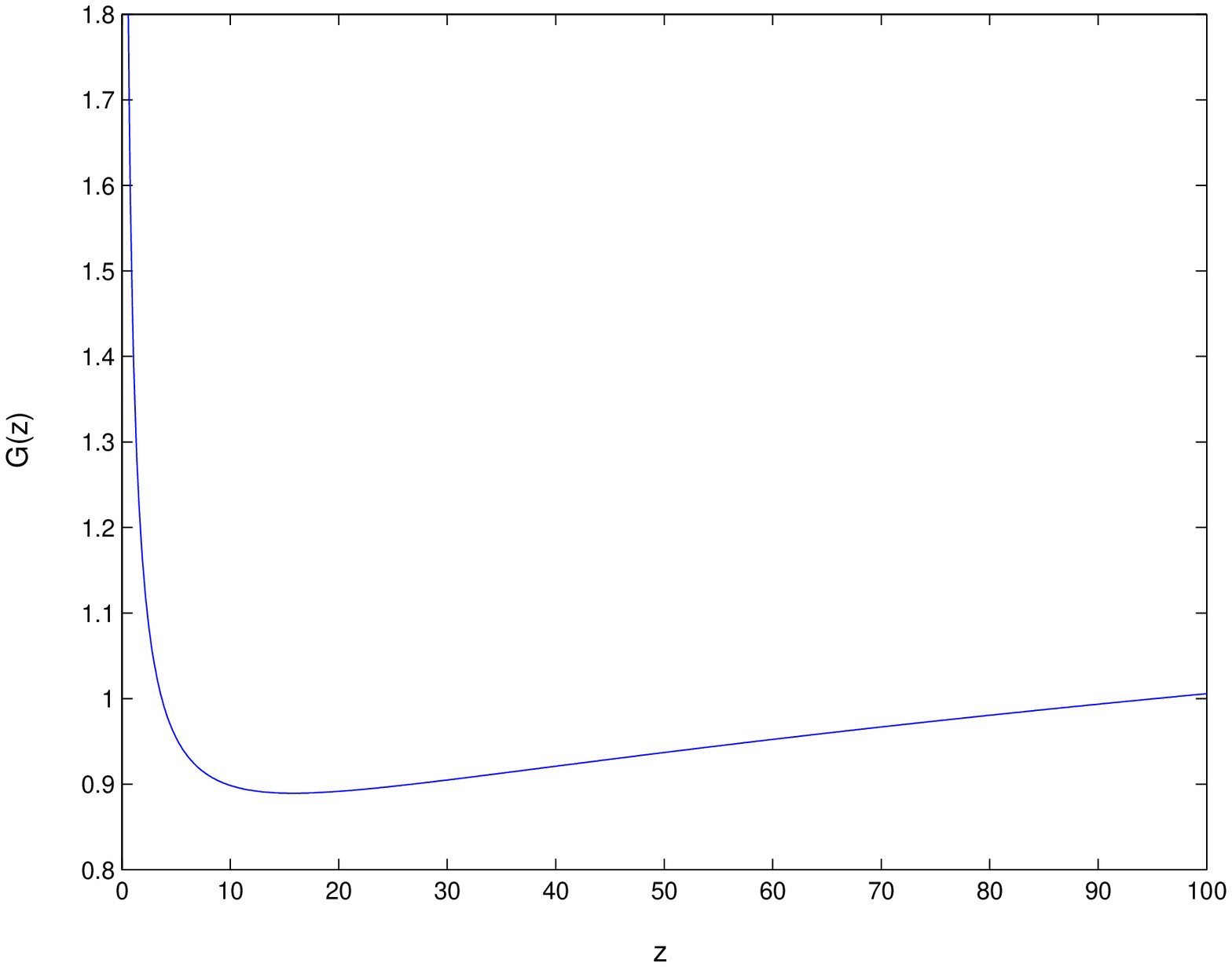}}
\caption{The gain function ${\cal G}(z)$ relating the radiation
intensity, frequency, and temperature.\label{fig1}}
\end{figure}

For a fixed and frequency-independent $I_\nu\equiv I$, very different
temperatures are found at different frequencies. We rewrite
Eq.~(\ref{temperature}) and (\ref{temperature/intensity}) in terms of the
reciprocal of the mean occupation number, $z\equiv h\nu^3/(c^2I_\nu) =
1/n$,
\begin{equation}
\label{tempmod}
k_BT_\nu = h\Big({{c^2 I} \over h}\Big)^{1\over 3} {\cal G}(z) ,
\end{equation}
where
\begin{equation}
\label{gain}
{\cal G}(z) = z^{1\over3}/\ln(1+ z) .
\end{equation}
The gain function ${\cal G}(z)$ is plotted in Fig.~\ref{fig1} and
determines how the intensity and frequency are related to
the temperature for radiation. 
${\cal G}(z)$ is singular at $z = 0$, with a minimum at approximately $z =
15.8$; it grows gradually, unbounded, with increasing
$z$.

Clearly ``hotter'' radiation sources for a given energy are either at
low or very high frequencies. For example, a laser in the 600\,nm range
is less ``hot,'' watt for watt, than an X-ray laser with wavelengths of
the order of Angstroms. This result makes sense because
Eq.~(\ref{temperature/intensity}) must give a constant value for a
black body, having a minimum in the middle of the frequency domain.
Alternatively, the curve represents points where the black body
distribution crosses a constant specific intensity $I$. It crosses at
two frequencies except when the temperature falls so low that it does
not cross at all.

The low- and high-frequency limits of Eq.~(\ref{tempmod}) can be
understood differently. If we count photons as particles in terms
of their energies $\epsilon = h\nu$, the relation
$n = [\exp (\epsilon/k_BT)-1]^{-1}$ is independent of $h$. In the limit
$\epsilon\rightarrow
\infty$ or $T\rightarrow 0$, $n\rightarrow\exp (-\epsilon /k_BT)$, the
Maxwell-Boltzmann distribution. Photons in this limit act as classical
particles, and $n$ tends to be small
$(z\rightarrow\infty)$. Then consider the low-frequency or high-intensity
limit of Eq.~(\ref{tempmod}). In terms of intensity, the resulting
classical relation
$k_BT_\nu = (c^2/\nu^2)I$ is independent of $h$ and arises from a set of
classical thermal oscillators. In this Rayleigh-Jeans limit,
$z\rightarrow 0$ and $n\rightarrow\infty$.

Photons therefore have two different classical limits, at high frequencies
as particles and at low frequencies as a classical field. The field limit
is possible because photons are bosons, and large numbers of photons can
coexist in the same field mode. Coherent states can be constructed as
analogues to classical field states.

\section{The Temperature of a Real Laser}
\label{reallaser}

Although lasers do not have infinite temperatures, even common, low-power
lasers have temperatures closer to those of stellar interiors than to everyday
matter temperatures, far exceeding the mistaken equipartition estimates of
Sec.~\ref{misapp}. Lasers operate far into the high-intensity Rayleigh-Jeans
limit of Eqs.~(\ref{temperature})--(\ref{temperature/intensity}), where
$k_BT_\nu = (c^2/\nu^2)I_\nu$. To see these very high temperatures, the
specific intensity needs to be extracted from the laser power $P$.

The flux density $F$ is related to the specific intensity $I_\nu({\bf\hat k})$
by
\begin{equation}
\label{laserEg}
F = dP/dA = \!\int\! d\nu\, d\Omega_{\bf k} ({\bf{\hat r} \cdot {\hat
k}}) I_\nu({\bf\hat k}) .
\end{equation}
The surface area element $dA = r^2d\Omega_{\bf
r}$, and $\cos\theta = {\bf{\hat r} \cdot {\hat k}}$ is the cosine of
the angle between the wavevector and surface area normal vector
${\bf{\hat r}}$. $d\Omega_{\bf k}$ is the mode solid angle element, and
$dA$ represents the differential exit aperture area of the laser beam.

If $I_\nu({\bf\hat k})$ were constant with beam angle and frequency, then
$I_\nu({\bf\hat k})= I$, and the power from each point of the aperture
would be constant over the aperture, $F = P/A$.  The power would be
\begin{equation}
P = \pi AI\sin^2\theta_{1/2},
\end{equation}
where $\theta_{1/2}$ is the beam spread half-angle, and $A$ is
the aperture area of the laser.


Let the specific intensity have the factorized form:
\begin{equation}
\label{factorized}
I_\nu({\bf\hat k}) = F_0 \Phi (\nu) {\cal D}({\bf\hat k}) .
\end{equation}
Each function, $\Phi (\nu)$ and ${\cal D}({\bf\hat k})$, is separately
normalized to unity:
\begin{equation}
\int^\infty_0\! d\nu\, \Phi (\nu) = \! \int\! d\Omega_{\bf k} ({\bf{\hat
r} \cdot {\hat k}}) {\cal D}({\bf\hat k}) = 1 .
\end{equation}
The solid angle variables are $\theta$ and $\varphi$. They are the polar
angle relative to, and the azimuthal angle about, the beam axis,
respectively. 

We will consider typical lineshapes and angular
distributions.\cite{sargent74,silf96}
The simplest lineshape is the Gaussian form arising from Doppler
broadening by thermal agitation of the lasing gas,
\begin{equation}
\label{spectralshape}
\Phi (\nu) = \exp [-(\nu
-\nu_0)^2/(\Delta\nu_D)^2]/\sqrt{\pi}\Delta\nu_D ,
\end{equation}
and $\Delta\nu_D$ is the half-width due to Doppler broadening:
\begin{equation}
\Delta\nu_D = \nu_0\sqrt{\frac{2k_BT_{\rm gas}}{mc^2}} ,
\end{equation}
$m$ and $T_{\rm gas}$ are the mass and temperature of the gas atoms
emitting the radiation.

The simplest angular distribution is also a Gaussian from the laser
cavity resonating in its fundamental mode.
\begin{subequations}
\label{angularshape}
\begin{eqnarray}
{\cal D}(\theta) = 2 \exp [-2\theta^2/\theta^2_0]/\pi\theta^2_0, \\
\int^{2\pi}_{0}\! d\varphi \!\int^{+1}_{-1}\! d(\cos\theta) \cos\theta
{\cal D}(\theta) = 1 .
\end{eqnarray}
\end{subequations}
The distribution is
azimuthally symmetric about the forward beam direction, with $\theta_0\ll
1$ as the half-angle,
$e^{-2}$-power, beam divergence. An ideal laser's beam divergence is
diffraction-limited at the aperture: $\theta_0 = 2\lambda_0/\pi D$,
where
$D$ equal to the aperture diameter.

\begin{figure}[h]
\centerline{\epsfxsize=7.0in \epsfbox{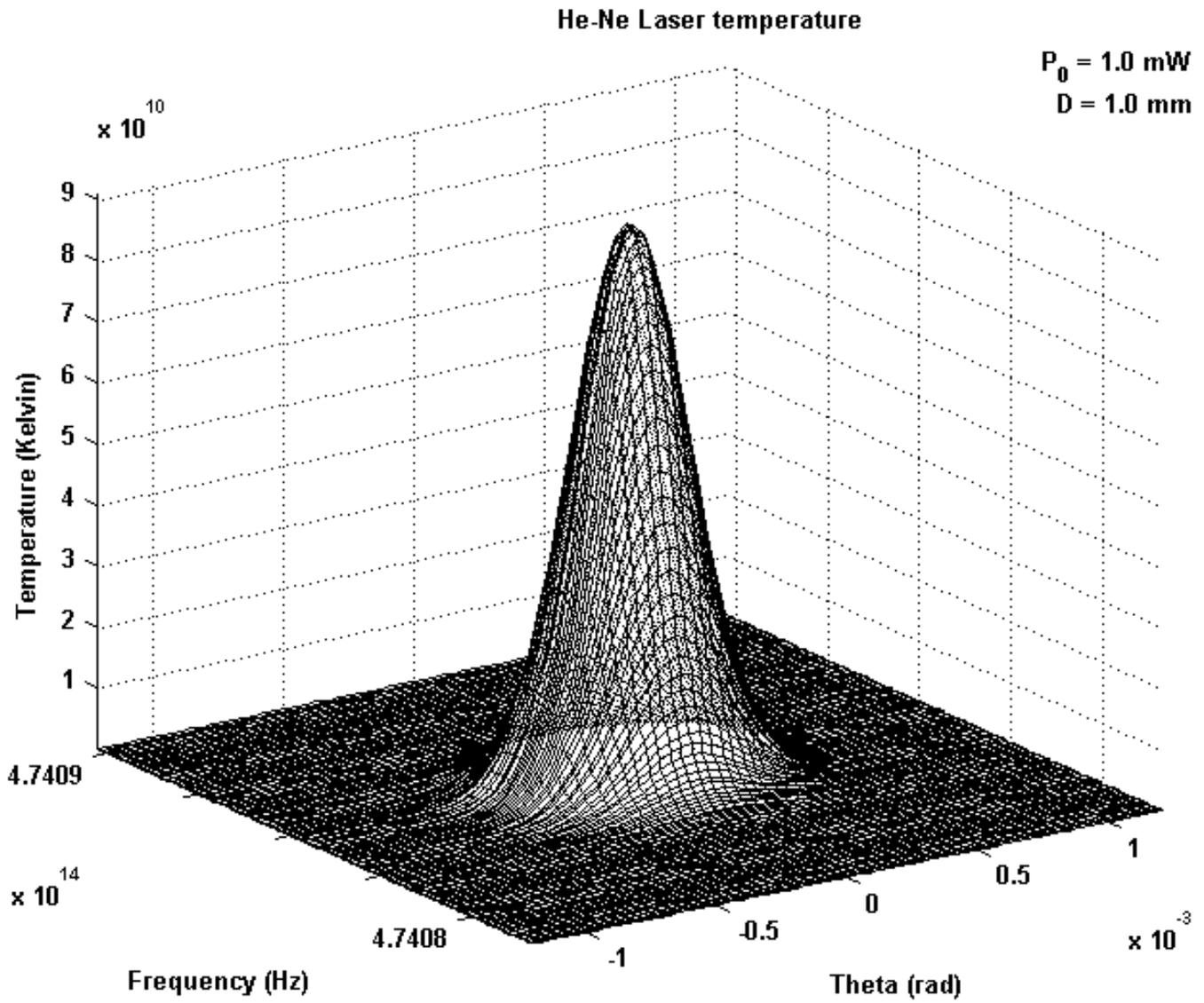}}
\caption{The helium-neon laser radiation temperature $T(\nu ,\theta)$
for $D = 1$\,mm, $\Delta\nu_D = 0.9$\,GHz, and $P_0 =
1.0$\,mW.\label{fig2}}
\end{figure}

Our results are for a red He-Ne gas laser, wavelength
$\lambda_0 = 632.8$\,nm or
$\nu_0 = 4.741\times 10^{14}$\,Hz. We assume a power of $P_0 =
1$\,mW. Figure~\ref{fig2} is a plot of the radiation temperature
$T(\nu,\theta)$ for $D = 1$\,mm and
$\Delta\nu_D = 0.9$\,GHz, corresponding to $T_{\rm gas}\simeq 390$\,K.
The beam aperture area $A$ is $\pi D^2/4\simeq 8\times 10^{-7}$\,m$^2$,
and the total flux density $F_0 = P_0/A\simeq 1.3\times 10^3$\,W/m$^2$.
The beam divergence $\theta_0\simeq 0.4$\,mrad.

\begin{figure}[h]
\centerline{\epsfxsize=6.8in \epsfbox{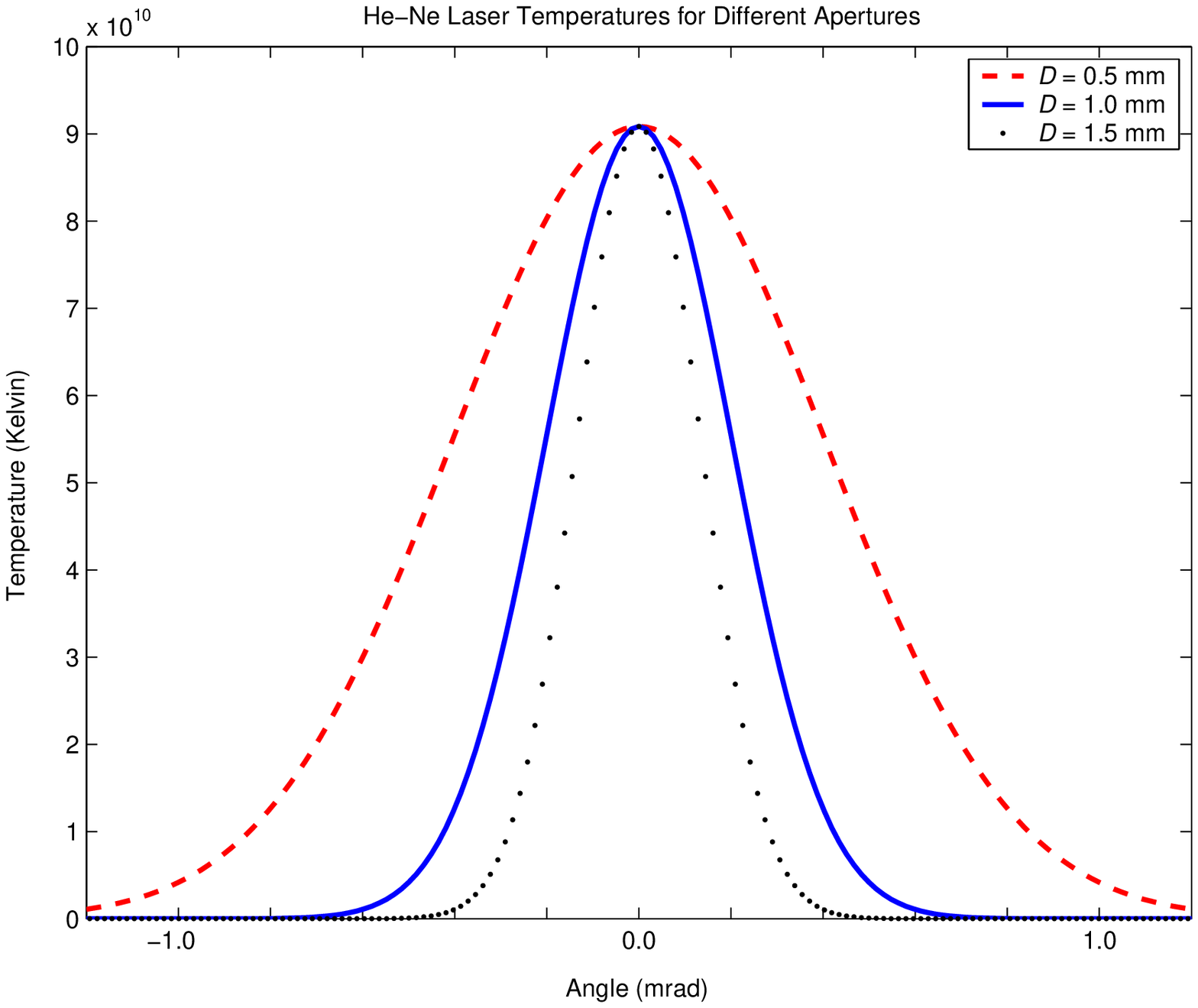}}
\caption{The helium-neon laser radiation temperature $T(\nu_0,\theta)$
for three aperture diameters $D$. The intensity/temperature profile
narrows as
$D$ increases and the beam becomes more parallel.\label{fig3}}
\end{figure}

Figure~\ref{fig3} shows the radiation temperature $T(\nu =\nu_0,\theta)$ for several beam
diameters $D$ and corresponding beam divergences $\theta_0$, with the same
$P_0$ and $\Delta\nu_D$. The peak temperature is
\begin{equation}
k_BT_{\nu}(\nu = \nu_0, \theta = 0) =
\frac{2c^2F_0}{\pi\sqrt{\pi}\nu^2_0\theta^2_0\Delta_D} .
\end{equation}
It is independent of $D$, because the normalization of temperature is
independent of $D$:
\begin{equation}
T\sim F_0/\theta^2_0\sim
P_0\theta^{-2}_0/A\sim \theta^{-2}_0D^{-2}\sim D^2 D^{-2} .
\end{equation}
Only the shape of the angular distribution depends on $D$.
Note that $T_{\nu}$ diverges if $\Delta_D$ or $\theta_0\rightarrow 0$,
reproducing the pure state of Sec.~\ref{lagrange}.


To infer the radiation temperature of a laser requires knowing the
specific intensity $I_{\nu}$ of its light. Because a laser's radiation is
so intense, we can use the Rayleigh-Jeans limit of
Eq.~(\ref{temperature/intensity}):
$k_BT_{\nu} =
(c^2/\nu^2)I_{\nu}$. Finding $I_{\nu}$ requires making separate geometry-
and frequency-based measurements. The former requires an intensity
measurement at a fixed frequency, typically with a photodetector. The
latter determines the spectral characteristics of the
beam.\cite{silf96,melles03,coherent03}

We place a photodetector of known area $A$ and frequency response or
measurement efficiency
$R(\nu)$ orthogonally
across the beam and centered along the beam axis $(\theta = 0$), at a
distance $r$ from the beam exit. $R(\nu)$ is normalized so that its
integral is unity. The photodetector measures an incoming power $P$:
\begin{equation}
\label{measuredpower}
P = 2\pi A \!\int^{\infty}_0\! d\nu\! \int_{A} d(\cos\theta )\cos\theta 
R(\nu) I_{\nu}(\nu,\theta) ,
\end{equation}
with the definition~(\ref{laserEg}). The factor of $2\pi$ represents
azimuthal integration around the beam axis and assumes axial symmetry
of the beam's intensity.

We make the following simplifying assumptions.

\begin{enumerate}
\item Assume that the frequency and angular profiles are Gaussian,
like the expressions (\ref{spectralshape}) and (\ref{angularshape}).
From these profiles, we can infer the peak flux density $F_0$ with a few
measurements.

\item The solid angle subtended by the photodetector is $\Omega = A/4\pi
r^2$. Assume the photodetector itself is circular. The half-angle
$\theta_A$ it subtends with the beam axis is given by $\tan\theta_A =
\sqrt{A/\pi r^2}$.

\item All relevant angles, $\theta_A$ and $\theta_0$, are small. Then
$\cos\theta \simeq
1 - \theta^2/2$. The angular integration becomes:
\begin{equation}
\int_{\cos\theta_A}^{1}\! d(\cos\theta) \cos\theta I_{\nu}(\nu ,\theta)
\doteq \!\int^{\theta_A}_0\! d\theta \theta I_{\nu}(\nu ,\theta) .
\end{equation}

\item The frequency integration can be simplified if we assume that
$R(\nu)$ is constant across a range $\Delta F$, $R(\nu) = (\Delta
F)^{-1}$. If we do the frequency integration numerically, however, this
simplification is not necessary.
\end{enumerate}

We can use values of $\nu_0$, $\Delta\nu_D$, and $\theta_0$ that are
specified by the laser's manufacturer. With additional instruments, we
can measure these quantities, although such measurements are more
difficult than the power measurement. For example, we can determine the
frequency profile and infer $\nu_0$ and $\Delta\nu_D$ by measuring the
specific intensity as a function of wavelength.  Light
wavelength is typically measured with a grating spectrometer. But a
grating does not have the frequency resolution needed for a laser, and
instead we should use a Fabry-Perot interferometer-based analyzer.
We can determine the angular profile and infer $\theta_0$ by
measuring how the relative intensity varies as we vary $r$, the
distance of the photodetector from the beam aperture. Because we move
the photodetector farther away from the laser, it subtends a larger
half-angle $\theta_A$ with the beam axis.

If we substitute these derived values and assumed simplifications
into Eq.~(\ref{measuredpower}), we can do the angular integral
analytically. The frequency integral can be done numerically. With the
measured $P$, $A$, $\nu_0$,
$\Delta\nu_D$, and $\theta_0$, we can infer $F_0$ and $I_{\nu}(\nu
,\theta)$ from Eqs.~(\ref{factorized}), (\ref{spectralshape}),
(\ref{angularshape}), and (\ref{measuredpower}). Your results should
be similar to Figs.~\ref{fig2} and \ref{fig3}, unless
your laser is not operating in its fundamental frequency mode.
(If the photodetector is calibrated in absolute power units, your value of
$F_0$ is also in absolute units. If you cannot measure $P$ in absolute
units, you can still measure a relative intensity profile in frequency and
angle, normalized to the peak intensity. The resulting temperature is
also in relative units.)

\section{Other Physical Consequences}
\label{otherphysconseq}

Nonequilibrium radiation temperature and its large magnitude in lasers have
important theoretical and practical consequences.

Consider the entropy production density of radiative transfer
(matter-radiation coupling). The matter interacting with the radiation has
a local temperature $T$. This entropy production density can be related
to the radiation energy specific intensity $I_\nu$ and a entropy specific
intensity $J_\nu$\cite{essex99}:
\begin{eqnarray}
\label{entropyproduction}
-{{\dot{u} +\nabla\cdot\mathbf{F}}\over{T}}&+& \dot{s}_r +
\nabla\cdot\mathbf{H}
= \! \int \! d\nu d\Omega_{\bf k} \big\{ -({\bf\hat{k}}\cdot
\nabla I_\nu +
{1 \over c}\dot{I}_\nu){1 \over T} + ({\bf\hat{k}}\cdot \nabla J_\nu +
{1 \over c}\dot{J}_\nu)\big\}
\nonumber \\ 
 & = & \! \int \! d\nu\ d\Omega_{\bf k} \bigg\{
\Big({\bf\hat{k}}\cdot
\nabla I_\nu + {1\over c}\dot{I}_\nu\Big)
\big({{\partial J_\nu}\over {\partial I_\nu}}- {1 \over T}\big) \bigg\}
\nonumber \\
 & = & \!\int\! d\nu d\Omega_{\bf k} \big\{
\Bigl({\bf\hat{k}}\cdot
\nabla I_\nu + {1\over c}\dot{I}_\nu\Bigr)
\big({1 \over T_\nu (\mathbf{\hat{k}})}- {1\over T}\big)\big\}.
\end{eqnarray}
$\mathbf{H}$ and $\mathbf{F}$ are vector densities of radiative entropy and energy
flux, respectively, and
$u$ and $s_r$ are radiative energy and entropy densities, respectively.
The radiation temperature
$T_\nu (\mathbf{\hat{k}})$ is defined in Eqs.~(\ref{tempderivative}),
(\ref{temperature/intensity}), and (\ref{temptransport}).
The first term on the first line is the matter entropy production density;
the second and third are the radiation entropy production density.

Equation~(\ref{entropyproduction}) is positive semidefinite and expresses
the second law of thermodynamics in radiative transfer. The left-hand
factor in the last integrand of Eq.~(\ref{entropyproduction}) is the energy
transfer rate density, which is equal and opposite for matter and
radiation.\cite{mihalas99} For strictly thermal emission, this
factor corresponds to the net cooling or heating of matter due to the
radiation beam in the direction ${\bf\hat{k}}$.

\begin{itemize}
\item If $T > T_\nu (\mathbf{\hat{k}})$, it is positive: the matter
cools, the field gains energy, and the entropy production is positive.

\item If the temperature inequality is reversed, the sign of this factor
also reverses: the field loses energy, the matter is heated, and the
entropy production remains positive.

\item The product in the integral is zero only when the matter and
radiation temperatures are equal and no transfer occurs.

\end{itemize}

The two other ways matter can transfer heat to other matter are by
convection and diffusion. These mechanisms require matter-matter
contact and gradients in intensive variables such as temperature.
Radiative energy transport, on the other hand, depends only on the
difference of the local matter and radiation temperatures at a single
point in space. Even a low-power laser is very effective at transferring
energy, because its radiation temperature is so large. Energy transfer
dominated by radiative processes is minimally affected by convection
and conduction if the matter temperature is much smaller than the
radiation temperature.

Practical consequences of these results include the following.
Because of the high beam temperature, radiative transfer dominates over
the other, more destructive and undesirable heat transfer processes of
convection and conduction, making lasers effective tools for surgery.
There is neither the time nor the energy to induce gradients large
enough to make convection or conduction
important. The overall deposited energy is small, and the cut tissue
suffers little damage. The cooling of atoms with lasers is also
interesting in light of the results of this paper.
In Sec.~\ref{infinite}, two cases of vanishing entropy were considered.
Case~2 represents the laser radiation cooling the atoms, while Case~1
conforms to the behavior of the atoms being cooled. When the atoms
decrease in entropy, according to Case~1, their temperature and energy
must also go down. But the radiation behaves according to Case~2: its
temperature also goes down, but its entropy increases instead, while
its power stays constant.

The classical field limit of $T_\nu ({\bf\hat k})$ is obtained by
letting $h\rightarrow 0$ in Eq.~(\ref{temperature/intensity}) and
holding everything else constant. The result, $k_BT_\nu ({\bf\hat k}) = 
h\nu n = c^2I_\nu ({\bf\hat k})/\nu^2$,
is identical to the high-intensity or low-frequency limit, independent of
$h$ if written in terms of $I_\nu ({\bf\hat k})$. The magnitude of the
flux density
$|\mathbf{F}| = F$ is
\begin{equation}
\label{radnoise}
F = \!\int\! d\nu d\Omega_{\bf k} k^2 ({\bf{\hat r}\cdot{\hat k}}) k_BT_\nu
({\bf\hat k}) .
\end{equation}
This result can be compared to the antenna or noise temperature
familiar in radio-frequency electromagnetism.\cite{christ96} In a
one-dimensional system, the power $P_{\rm noise}$ of a pure noise signal
is associated with a temperature
$T_{\rm noise}$:
\begin{equation}
\label{elecnoise}
P_{\rm noise} = \!\int^{\infty}_0\! d\nu k_BT_{\rm noise}(\nu) .
\end{equation}
The result~(\ref{radnoise}) is the three-dimensional, solid-geometry
analogue of the radio-frequency antenna or circuit
result~(\ref{elecnoise}). It should be stressed that the antenna
temperature is the temperature of the radiation, not the antenna
material; just as $T_\nu({\bf\hat k})$ is not the temperature of the
lasing gas, but of the laser light.

The arguments of Sec.~\ref{radthermo} make clear that the noise
temperature defined in Eqs.~(\ref{radnoise}) and (\ref{elecnoise})
requires a randomized phase distribution for the noise signal, in
keeping with the usual intuitive definition of noise.

The results of this paper demonstrate some of the physical implications of
temperature for nonequilibrium systems, in particular for radiation and laser
light.  More consequences can be inferred by applying the techniques and results
presented here to other aspects of radiation, radiation transport, and lasers,
and can be explored starting with the references.

\begin{acknowledgments}

The authors acknowledge the hospitality and stimulus of the Telluride
Summer Research Center. R.\ S.\ B.\ acknowledges the support of the
National Science Foundation. We are grateful as well to David Reitze
(University of Florida) for explaining laser measurements and to Sidney
Bludman (University of Pennsylvania and DESY) for helpful comments. The
referees improved the form and content of the paper. The figures were
generated using MATLAB 6.

\end{acknowledgments}

\newpage

\begin{thebibliography}{99}

\bibitem{sargent74} M. Sargent III, M. O. Scully, and W. E. Lamb, Jr.,
{\it Laser Physics} (Addison-Wesley, Reading, MA, 1974),
Chaps. 6--8 and 14--17.

\bibitem{silf96} W. T. Silfvast, {\it Laser Fundamentals} (Cambridge
University Press, 1996), Chap. 6 and Parts III, IV, and V.

\bibitem{mandel95} L. Mandel and E. Wolf, {\it Optical Coherence and
Quantum Optics} (Cambridge University Press, 1995), Chaps. 10--13, 15,
and 18.

\bibitem{essex99} C. Essex and D. C. Kennedy, Minimum Entropy Production
of Neutrino Radiation in the Steady State, J. Stat. Phys. {\bf 94},
253--267 (1999).

\bibitem{reichl98} L. Reichl, {\it A Modern Course in Statistical
Physics} (John Wiley \& Sons, 1998), 2nd ed., Chaps. 2, 6, and 10, and
Appendix B.

\bibitem{sien93} S. Sieniutycz and R. S. Berry, Canonical formalism, fundamental
equation, and generalized thermomechanics for irreversible fluids with heat transfer,
{\it Phys. Rev.} E {\bf 47}, 1765--1783 (1993).

\bibitem{degroot62} S. R. de Groot and P. Mazur, {\it Non-Equilibrium
Thermodynamics} (Dover Publications, 1984), Chaps. 1, 3--5.

\bibitem{bludman97} S. A. Bludman and D. C. Kennedy, Variational Principles
for Stellar Structure, Astrophys. J. {\bf 484}, 329--340 (1997).

\bibitem{merz69} E. Merzbacher, {\it Quantum Mechanics} (John
Wiley \& Sons, 1969), 2nd ed., Chaps. 15 and 20--22.

\bibitem{mihalas99} D. Mihalas and B. Weibel-Mihalas, {\it Foundations
of Radiation Hydrodynamics} (Dover Publications, 1999), Chap. 6.


\bibitem{fnote1} A stream of laser
radiation is almost coherent, but the initial choice of the radiation
phase is random. Once chosen, the phases of all the laser atoms are
slaved to it. For simplicity, we ignore the major exception,
mode-locked lasers, in which external signals and internal nonlinear
mode couplings pick and lock a phase for the radiation mode itself.
Without such locking, a sharply-defined initial radiation phase
statistically diffuses until its ensemble is random over the interval
$\phi\in [0, 2\pi)$.

\bibitem{fnote2} The most general coherent state representation of
$\hatrho$ is $\int \! d^2\alpha\ d^2\beta R(\alpha ,\beta)
|\alpha\rangle\langle\beta |$, but this form is not necessary for our
purposes.\cite{sargent74}

\bibitem{fnote3} This result reflects the complementarity of photon number
and field phase. The quantum phase (Susskind-Glogower) operator
$\exp(i{\hat\phi})\equiv {\hat a}/\sqrt{{\hat a}{\hat a}^{\dagger}} =
\sum_n |n\rangle\langle n+1|$ is nondiagonal in the $|n\rangle$ basis.
The commutator $[\exp(i{\hat\phi}), {\hat n}] = \exp(i{\hat\phi})\neq
0$. See Ref.~\onlinecite{walls94} for a detailed exploration of
radiation phase in the quantum case.

\bibitem{walls94} D. F. Walls and G. J. Milburn, {\it Quantum Optics}
(Springer-Verlag, 1994), Chap. 2.

\bibitem{fnote4} For many laser applications, it is reasonable to
approximate the energy density of the beam as $u = h\nu n = I/c$. But
this relationship implies an exactly parallel beam and an unphysical
intensity delta function in the beam angle.

\bibitem{melles03} See ``Laser beam measurement'' at {\tt
{<}http://www.mellesgriot.com/products/lasers/{>}}

\bibitem{coherent03} See ``Lasers'' at {\tt
{<}http://www.coherentinc.com/Products/{>}}.

\bibitem{christ96} D. Christiansen, {\it Electronic Engineer's Handbook}
(McGraw-Hill, 1996), 4th ed., Secs. 17.1.3 and 22.1.14.

\end{thebibliography}
\end{document}